\documentclass{XrU2005}

\usepackage{graphicx}
\usepackage{natbib}

\begin{document}

\title{An End-to-End Test of Neutron Stars as Particle Accelerators}
\author{Patrizia A. Caraveo}
\affil{INAF IASF-Milano, Via Bassini, 15; 20133 Milano; Italy}


\maketitle

\begin{abstract}
Combining resolved spectroscopy with deep imaging, XMM-Newton is
providing new insights on the particle acceleration processes
long known to be at work in the magnetospheres of isolated
neutron stars. According to a standard theoretical interpretation,
in neutron stars' magnetospheres particles are accelerated along
the B field lines and, depending on their charge, they can either
move outward, to propagate in space, or be funnelled back,
towards the star surface. While particles impinging on the
neutron star surface should heat it at well defined spots,
outgoing ones could radiate extended features in the neutron star
surroundings. \\By detecting hot spots, seen to come in and out of
sight as the star rotates,  as well as extended features trailing
neutron stars as they move in the interstellar medium, XMM-Newton
provides the first end-to-end test  to the particle acceleration
process.

\end{abstract}

\section{Introduction}

Isolated neutron stars (INSs) are natural particle accelerators.
Their, presumably dipolar, rapidly rotating magnetic fields,
naturally inclined with respect to the star rotation axis, induce
electric fields ideally suited to accelerate particles already
present in the stars' magnetospheres or extracted from the crusts.
Following the seminal paper of Goldreich and Julian (1969) and
Sturrock (1971), a lot has been done to work out the details of
such an acceleration, focusing on its most likely location(s)
inside the INS magnetosphere and on its efficiency.
Traditionally, two classes of models have been developed: on one
side the  polar cap ones(Ruderman \& Sutherland, 1975, Harding \&
Daugherty, 1998, Rudak \& Dyck, 1999), where the acceleration
takes place near the star surface, just above the magnetic pole;
on the other side, the outer gap ones(Romani, 1996), where the
acceleration is taking place in the outer magnetosphere, not far
from the light cylinder. Recently, the slot gap model, extending
from the polar cap to the light cylinder, has been added as a
third alternative (Muslinov \& Harding, 2003, Dyck \& Rudak, 2003,
Harding, 2005). Notwhistanding  important differences between
models, the interaction between  accelerated particles (typically
electrons) and the star magnetic field results in the production
of high energy gamma-rays which, in turn, are not able to escape
the highly magnetic environment and are converted into electron
positron pairs. This initiates a cascade  rapidly filling the
magnetosphere with energetic particles which, interacting with
the magnetic field, are responsible for the vast majority of the
INSs' multiwavelength phenomenology.

\section{Acceleration by-products}

INSs are mainly studied through their non thermal radio emission.
Radio searches have been highly successful and the current radio
catalogues list more than 1500 pulsars (Manchester et al. 2005).
\\In spite of the sheer number of objects and their very diverse
phenomenology, INS radio emission accounts for a negligible
fraction of the star rotational energy loss. A far more important
fraction of the star energy reservoir goes into high-energy
radiation, mainly in high-energy gamma-rays. While the number of
objects shrinks to less than 1\% of the radio ones (Thompson et
al., 2001), in gamma rays the INSs' luminosity can reach a sizable
fraction of the total rotational energy loss, with an increase in
efficiency for the older objects.

The rich INSs' phenomenology encompasses now also X and optical
emissions. While the numbers are slightly higher than the
gamma-ray ones (a dozen in the optical (Caraveo, 2000, Mignani et
al. 2004) and two scores in X-rays (Becker \& Aschenbach, 2002)
the nature of the radiation is not as clear cut as in the radio or
gamma-ray domains. In the optical, as well as in X-rays, aged
neutron stars exhibit both thermal and non-thermal emissions.
Indeed, when non-thermal emission  somewhat weakens with age, the
thermal one begins to emerge to tell the story of the cooling
crust of the neutron star. \\INS thermal emission, however, is not
totally unrelated to the magnetospheric particle acceleration.
Depending on their electric charge, particles move in different
directions along the magnetic field lines. While those moving
outward try to escape the INS magnetosphere, those moving inward
hit the star and heat its crust at well defined spots, that,
under the assumption of a dipolar magnetic field, should coincide
with its polar caps (return currents, see e.g. Ruderman \&
Sutherland 1975; Arons \& Scharlemann 1979). Thus, thermal
emission could be of use to trace non-thermal phenomena.

The escaping particles, on the other side, are part of the
neutron stars' relativistic wind which is supposed to account for
the bulk of their observed rotational energy loss.
\\Such relativistic wind can be
traced through its interaction with the interstellar medium (ISM),
both in the immediate surroundings of the stars, where the INS
magnetic field is still important, or farther away, where the
wind radiation  pressure is counterbalanced by the shocked ISM.
An important player to determine the shape and the phenomenology
of the resulting Pulsar Wind Nebula (PWN) is the actual neutron
star speed. INSs are known to be high velocity objects and,
plunging supersonically  through  the ISM, they can give rise to a
rich bow shocks phenomenology seen in the radio, optical and
X-ray domains (e.g. Chattarjee \& Cordes, 2002).

Fig.1 summarizes the neutron star {\it acceleration tree}:
starting from the synoptic view (Harding, 2005) of the mechanisms
responsible the gamma ray emission (the aspect most intimately
related to the actual particle acceleration) and following the
destiny of the particle moving inward (left) and outward (right).
Using past and present space observatories operating at X and
gamma-ray domain, we can construct such a tree with the aim to
improve our understanding of the neutron star physics.

\begin{figure*}
\centering 
\includegraphics[width=\linewidth]{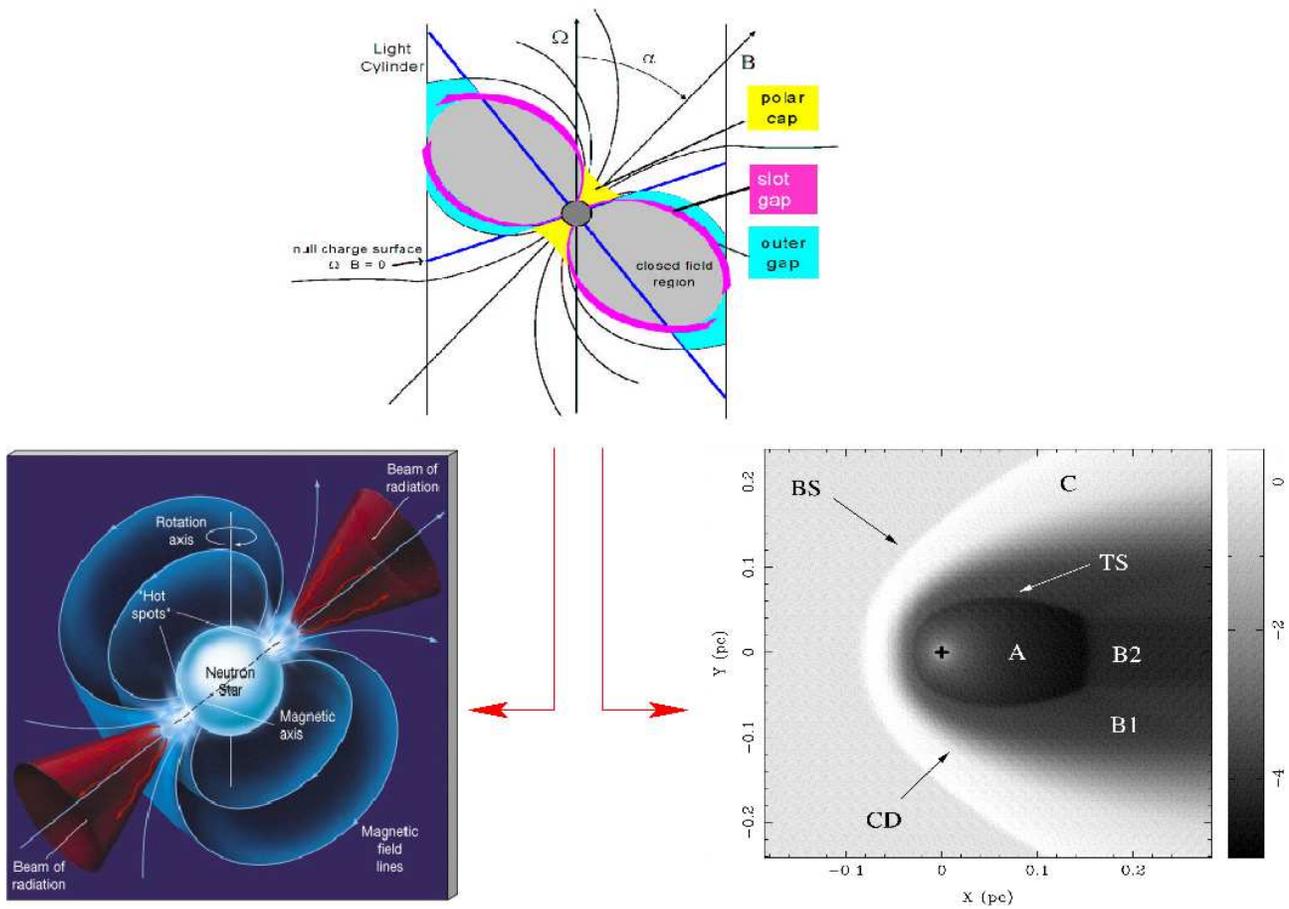}
\caption{Neutron star {\it acceleration tree}. Top panel :
schematic view of a pulsar magnetosphere showing the gamma-ray
emitting regions, according to the various classes of models
(from Harding, 2005). Bottom left: a similar view of a pulsar
magnetosphere showing the hot spots on the pulsar polar caps.
Bottom left: hydrodynamic simulation of a bow shock (BS)
generated by the interaction of the isotropic relativist wind of
a neutron star (marked with a cross), moving horizontally from
right to left, with the ISM (Gaensler et al, 2004). A indicates
the pulsar wind cavity, where the electrons propagate freely, B is
used for the shocked pulsar wind material, while C represents the
shocked ISM. The termination shock, TS, is where the energy
density of the pulsar wind is balanced by the external pressure,
while CD is the contact discontinuity bounding the shocked pulsar
wind material.}
\end{figure*}

\section{The observational panorama}

Our task is now to briefly review the data related to the three
steps highlighted in Fig.1 in order to find INSs observed
throughout the {\it acceleration tree}.

\subsection{High-energy gamma-rays}
Waiting for the next generation of gamma-ray instruments such as
Agile (Tavani et al., 2003) and Glast (Michelson et al.,2003), we
try to make the best use of the EGRET results. INSs (be they radio
loud or radio quiet) are the only class of galactic sources firmly
identified as high-energy gamma-rays emitters. Indeed, pulsars are
especially appealing to gamma-ray astronomers since their timing
signature allows to overcome the identification problem due to
the relatively large gamma-ray error boxes. However, in spite of
more than a decade of relentless efforts based on the EGRET data,
only Crab, Vela, PSR 1706-44, PSR1951+32, PSR1055-52 and the
radio quiet Geminga are confirmed gamma-ray sources (Thompson et
al., 2001). A few more pulsars have been proposed as sources of
pulsed gamma-radiation, but the claims are still awaiting
confirmations. Several positional coincidences between newly
discovered pulsars and old Egret sources (Hartman et al. 1999)
have been reported, but, again, such suggestions cannot be
confirmed without an operating g
amma-ray telescope. Thus, for the
moment being, our gamma-ray sample encompasses a very young (and
energetic) object such as the Crab, two slightly older pulsars
(Vela, PSR 1706-44) and three middle-aged INSs (PSR1055-57,
Geminga and the fast spinning PSR1951+32). It is worth noting
that the  efficiency for conversion of rotational energy loss into
gamma-rays changes as a function of pulsar age. It goes  from the
value of $\sim 0.009 \%$, for the Crab, to several \%, for Geminga
and PSR 1055-57.

\subsection{Hot spots}
The presence of hot spots on the surface of INSs has been long
suspected on the basis of their overall X-ray spectral shape
requiring more than a simple black-body to describe the data.
Usually two black-body curves, characterized by different
temperatures and emitting areas, are needed to fit the X-ray
spectra for all but the very youngest INSs. A slightly colder
black-body, covering the majority of the INS surface, provides the
bulk of the X-ray luminosity while a hotter one, covering a
smaller surface, is needed to obtain a satisfactory spectral fit.
\\Long XMM-Newton observations of Geminga, PSR0656+14 and PSR
1055-57, three middle-aged, rather similar INSs, have added an
important piece of information. Taking advantage of their
exceptional photon harvest, De Luca et al (2005a) were able to
perform  space resolved spectroscopy of the three INSs. For all
objects they have shown that  \\a) the spectra are varying
significantly throughout the rotational phase \\b)  the hot
blackbody contribution is the most dramatically variable spectral
component. \\This is shown in Fig.2 where the emitting radii,
computed on the basis of the phase-resolved spectral fits, are
shown as a function of the pulsar rotational phase. While for PSR
B0656+14 the modulation in the emitting radius wrt. the average
value is $<$10\%, similar to the value found for the cool
blackbody component, in the case of PSR B1055-52 we see a 100\%
modulation, since the hot blackbody component is not seen in 4
out of 10 phase intervals. A similar, 100\% modulation is
observed also for Geminga, although in this case the hot blackbody
component is seen to disappear in just one phase interval, and
the profile of its phase evolution is markedly broader.  It is
natural to interpret such marked variations as an effect of the
star rotation, which alternatively brings into view or hides one
or more hot spots on the star surface. As outlined above, such hot
spots arise when charged particles, accelerated in the
magnetosphere, fall back to the polar caps along magnetic field
lines. Straight estimates of neutron star polar cap sizes, based
on a simple ``centered'' dipole magnetic field geometry (polar
cap radius R$_{PC}=R\sqrt{\frac{R \Omega}{c}}$), where R is the
neutron star radius, $\Omega$ is the angular frequency and c is
the speed of light), predict very similar radii for the three
neutron stars, characterized by similar periods (233 m for PSR
B0656+14, 326 m for PSR B1055-52 and 297 m for Geminga, assuming
a standard neutron star radius of 10 km). The observed radii are
instead markedly different, with values ranging from $\sim 60$ m
for Geminga to $\sim$2 km for PSR B0656+14 (see De Luca et al,
2005a for a detailed discussion).

While waiting to enlarge the sample of deeply scrutinized X-ray
pulsars, it does not come as a surprise that two of the three
objects showing direct evidence for the presence of rotating hot
spots are highly efficient gamma-ray sources.

\begin{figure*}
\centering 
\includegraphics[width=\linewidth]{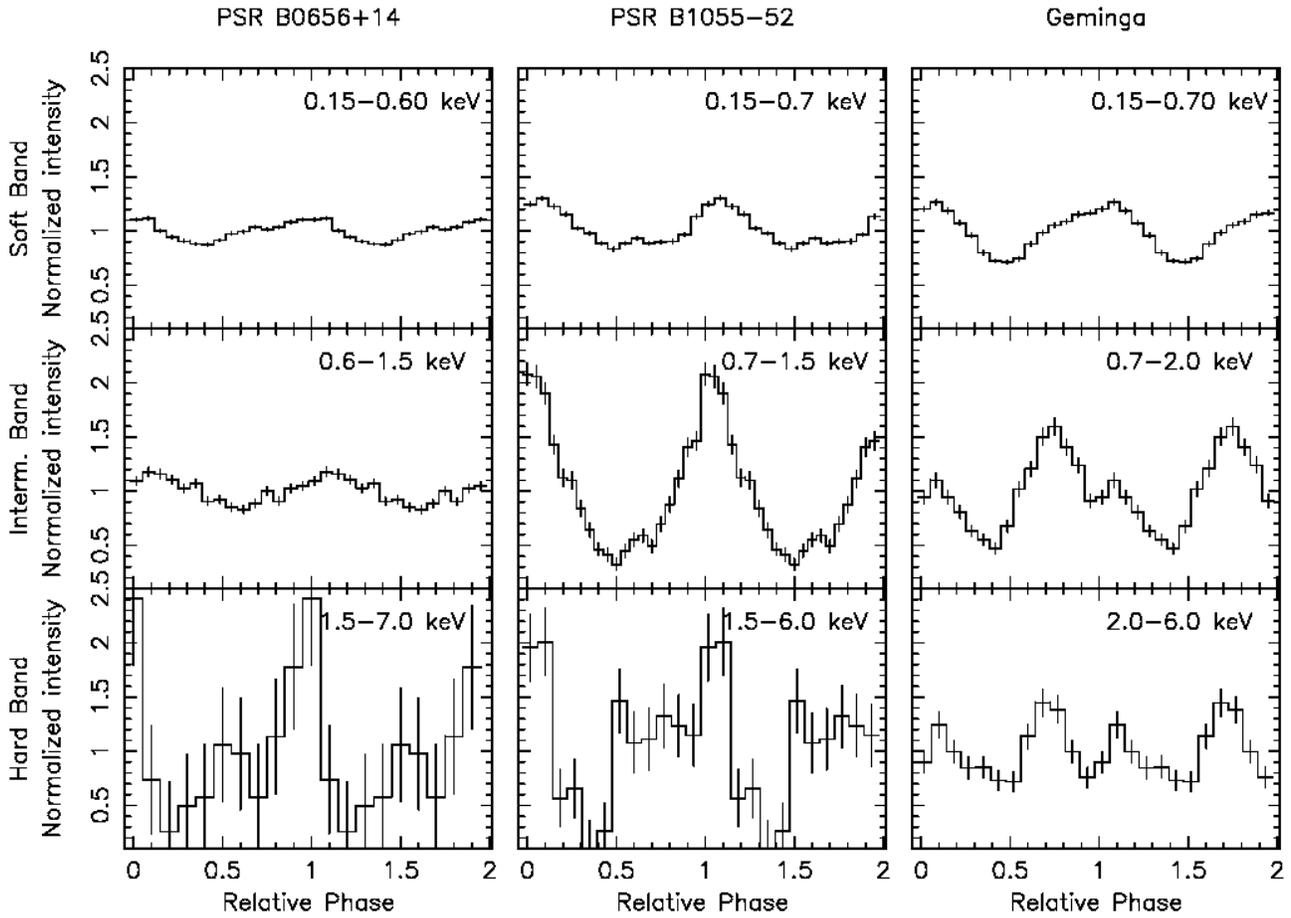}
\caption{Energy-resolved light curves of PSR B0656+14, PSR B1055-52 and Geminga in different energy ranges.
To ease the comparison of the behaviour of the three INSs, all light curves have been plotted setting phase 0
to the X-ray maximum. Pulsed fractions (computed as the ratio between the counts above the
minimum and the total number of counts) are as follows: PSR B0656+14 12.3$\pm$0.4\% in 0.15-0.6 keV,
16.9$\pm$2.3\% in
0.6-1.5 keV, 75$\pm$20\% in 1.5-7.0 keV; PSR B1055-52 16.7$\pm$0.6\% in 0.15-0.7 keV, 67$\pm$3\% in 0.7-1.5 keV,
90$\pm$10\% in 1.5-6.0 keV; Geminga 28.4$\pm$0.6\% in 0.15-0.7 keV, 54.5$\pm$2.4\% in 0.7-2.0 keV, 33$\pm$5\% in 2.0-6.0 keV.}
\end{figure*}

\subsection{Pulsar Wind Nebulae}
When the particle wind from a fast moving INS interacts with the
surrounding ISM, it gives rise to complex structures, globally
named ``Pulsar Wind Nebulae'' (PWNe) where $\sim10^{-5}-10^{-3}$
of the NS $\dot{E}_{rot}$ is converted into electromagnetic
radiation (for recent reviews see Gaensler et al 2004, and
Gaensler 2005). The study of PWNe may therefore give insights into
aspects of the neutron star physics which would be otherwise very
difficult to access, such as the geometry and energetics of the
particle wind and, ultimately, the configuration of the INS
magnetosphere and the mechanisms of particle acceleration.
Moreover, PWNe may probe the surrounding medium, allowing one to
measure its density and its ionisation state.

A basic classification of PWNe rests on the nature of the
external pressure confining the neutron star wind (e.g. Pellizzoni
et al. 2005). For young NSs ($<$ few $10^4$ y) the pressure
of the surrounding supernova ejecta is effective and a ``static
PWN'' is formed. For older systems ($>10^5$ y) the neutron
star, after escaping the eventually faded supernova remnant,
moves through the unperturbed ISM and the wind is confined by ram
pressure to form a ``Bow-shock'' PWN.

Static PWNe (Slane, 2005, for a review)  usually show complex
morphologies. Striking features such as tori and/or jets (as in
the Crab and Vela cases), typically seen in X-rays, reflect
anisotropies of the particle wind emitted by the energetic,
central INS and provide important constrains on the geometry of
the system. A remarkable axial symmetry, observed in several
cases, is assumed to trace the rotational axis of the central INS.
For the Crab and Vela PWNe, such an axis of symmetry was found to
be coincident with the accurately measured direction of the INS
proper motion (Caraveo \& Mignani 1999, Caraveo et al. 2001). This
provided  evidence for an alignement between the rotational axis
and the proper motion of the two neutron stars, with possible
important implications for the understanding of supernova
explosion mechanisms (Lai et al. 2001). The alignement between
spin axis and space velocity, directly observed only for Crab and
Vela, is now assumed as a standard property of NSs (Ng \& Roman,
2004).

Bow-shocks (for a review see Pellizzoni et al. 2005, Gaensler et
al 2004) have a remarkably simpler, ``velocity-driven''
morphology. They are seen frequently in $H_{\alpha}$ as arc-shaped
structures tracing the forward shock, where the neutral ISM is
suddenly excited. In other cases, X-ray emission (and/or radio
emission on larger scales) is seen, with a cometary shape
elongated behind the neutron star, due to synchrotron radiation
from the shocked NS particles downstream (only in the case of PSR
B1957+20 both the $H_{\alpha}$ and the X-ray structures have been
observed, Stappers et al., 2003). According to the lower
energetics of the central, older INS, bow shocks are typically
fainter than static PWNe and proximity is a key parameter for
their observation.
\\Since we aim at tracing the high energy particle escaping the
INS magnetosphere, we concentrate on the X-ray PWNe. Inspecting
the list of Gaensler et al (2004), we find only PSR B1951+32 in
common with the gamma-ray database, leaving little hope to find
an object displaying all the aspects of the  {\it acceleration
tree}. \\However, recent observations of Geminga,  combined with
previous ones by XMM-Newton, have unveiled the presence of a bona
fide PWN with complex diffuse features trailing the pulsar
perfectly aligned with its well known proper motion (De Luca et
al., 2005b; Caraveo et al. 2003).

Thus, the combined EGRET, XMM-Newton and Chandra results on
Geminga make this source the most suitable example  for our
end-to-end test of particle acceleration. For a review on the
multiwavelength phenomenology of Geminga, see Bignami \& Caraveo
(1996).

\begin{figure*}
\centering 
\includegraphics[width=\linewidth]{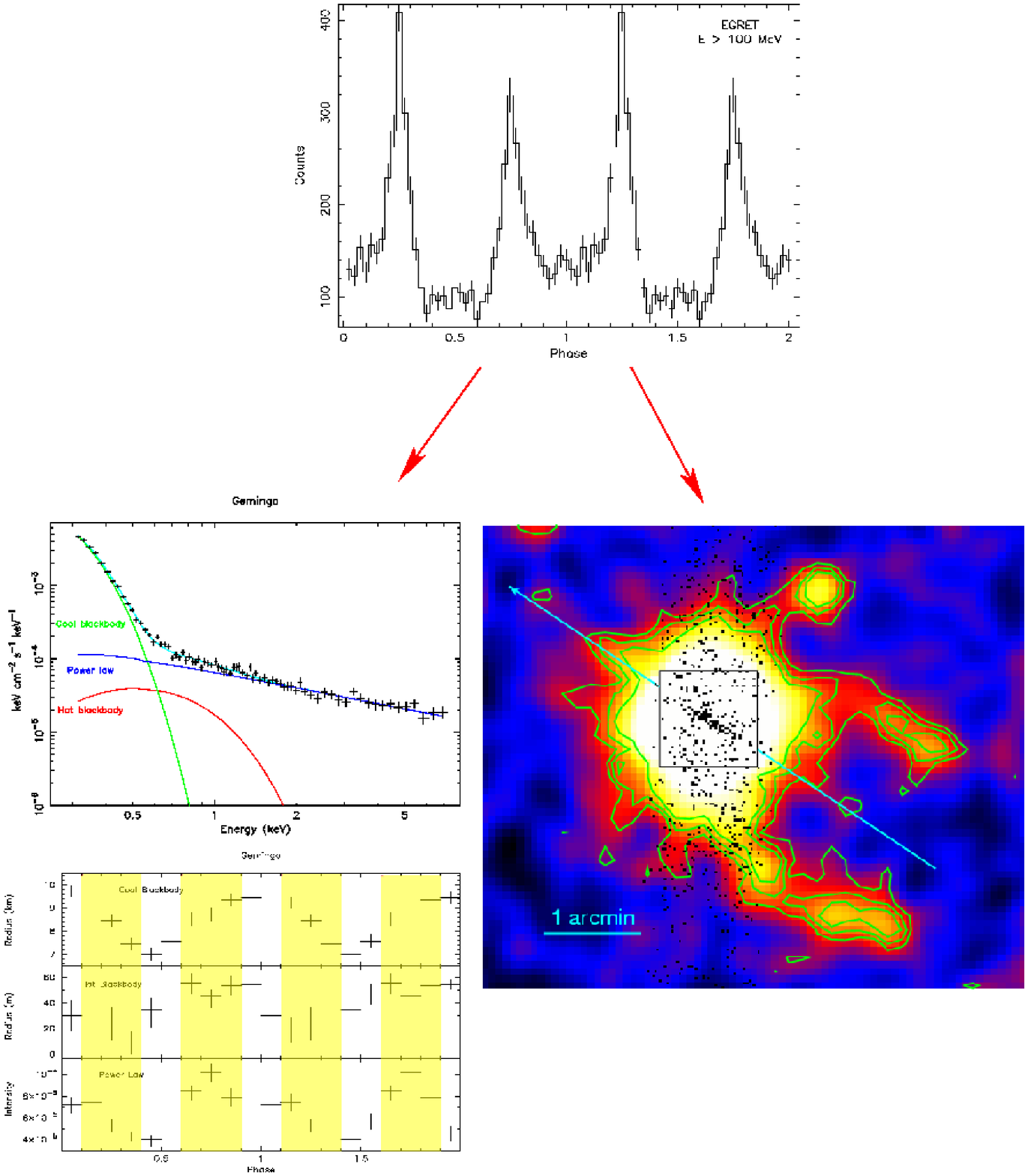}
\caption{The {\it acceleration tree} as
observed for Geminga. Top: the gamma-ray light curve. Left: the
XMM-Newton average spectrum as well as the results of
phase-resolved spectroscopy, showing the evolution of
the black-body emitting regions as a function of the INS
rotational phase. The shaded area mark the phase intervals
corresponding to the $\gamma$-ray peaks observed by EGRET. The
highest $\gamma$-ray peak occurs at phase $0.25\pm0.15$, the
second one at phase $0.75\pm0.15$ (the uncertainty is due to the
extrapolation of the EGRET ephemeris to the epoch of the
XMM-Newton observation).
Right: Geminga as seen by Chandra and
XMM-Newton (from De Luca et al., 2005b). The Chandra image,
rebinned to a pixel size of 2" has been superposed on the
XMM-Newton/MOS image obtained by Caraveo et al.(2003). Surface
brightness contours for the XMM image have been also plotted. The
ACIS field of view is limited to a rectangular region 1 arcmin wide.
The
pulsar proper motion direction is marked by an arrow.}
\end{figure*}

\section{Geminga as a test case}

Fig.3 summarizes all the observational evidence collected so far
on the presence of high energy  electrons/positrons in the
magnetosphere of Geminga. First, the EGRET light-curve whose $>$
100 MeV photons could not have been produced without high energy
particles and magnetic fields. \\Next, the contribution of a 100
ksec XMM-Newton observation which yielded both \\a) the evidence
for the presence of minute hot spot(s) varying throughout the
pulsar phase (Caraveo et al., 2004)
\\b) the detection of two elongated tails, trailing the pulsar in
its supersonic motion through the ISM and perfectly aligned with
the proper motion direction. The flat spectral shape of the
tails' X-ray photons suggests a synchrotron origin which, combined
with the typical magnetic field present in a shocked ISM, implies
the presence of $\sim10^{14}$ eV electrons/positrons, i.e. of
particle at the upper limit of the energy range achievable for an
INS like Geminga. Moreover, the lifetime of such electrons (or,
more precisely, the time it takes for them to lose half of their
energy) in the bow-shock magnetic field is $\sim800$ years.  On
the other hand, Geminga's proper motion (170 mas/year) allows one
to compute the time taken by the pulsar and its bow shock to
transit over the apparent length of the x-ray structures in the
sky (~3' from the central source). Such a time is close to 1,000
years. Thus, Geminga's tails remain visible for a time comparable
to the electron synchrotron X-ray emission life time after the
pulsar passage. The comet-like structure seen by Chandra is as
luminous as the larger and fainter tails and its spectrum is
equally hard.

Hot spot(s), elongated, faint tails and short, brighter trail
have roughly the same luminosity, corresponding to $\sim10^{-6}$
of its $\dot{E}_{rot}$.

We note that the morphology and hard spectrum of the Trail is
reminescent of the jet-like collimated outflows structures seen
in the cases of Crab and Vela (Helfand et al., 2001, Pavlov et
al. 2003, Willingale et al., 2001, Mori et al. 2004) and
associated to the neutron stars spin axis direction. In
particular, the small Geminga's Trail can be compared to the
``inner counterjet'' of the Vela PSR (Pavlov et a., 2003),
characterized by a similar spectrum (photon index $\sim$1.2) and
efficiency ($L_{X}$$\sim$$10^{-6} \dot{E}$). The projected angle
between Geminga proper motion and its backward jet is virtually
null, which implies that also the pulsar spin axis should be
nearly aligned with them. Geminga would thus be the third
observed neutron star having its rotational axis aligned with its
space velocity, after the cases of the Crab and Vela.\\The whole
scenario, encompassing both the large Tails and the small Trail,
could therefore fit in the frame of an anisotropic wind geometry.
It includes jet structures along the spin axis and relativistic
shocks in the direction of the magnetic axis where most of the
wind pressure is concentrated due to the near radial outflow from
magnetosphere open zones. \\The coupling of the jet-like Trail
seen by Chandra with the larger, arc-shaped Tails seen by XMM has
no similarity with other pulsars.

\section{Conclusion}
The particle acceleration going on in an INS magnetosphere can
now be traced from end-to-end. While gamma ray emission probes
directly the particle population in the magnetosphere, using the
current generation of X-ray observatories we are now able to
follow the destiny of the particles traveling up and down the
magnetic field lines through the study of  hot spots on the star
surface and of PWNe. The same process responsible for the copious
gamma-ray emission of Geminga would thus also be responsible for
the appearance of the hot spots on its surface (Halpern \&
Ruderman, 1993). Such a strong link between the X-and gamma-ray
behaviour of the source could be exploited to map the relative
positions of the regions responsible for the different emissions.
A precise comparison of the source X and gamma-ray light curves
is crucial at this point, but the lack of operating high energy
gamma ray telescope makes it impossible. Simultaneous observations
performed by XMM-Newton and by Agile and/or GLAST (foreseen to be
operational in the coming years) will add important pieces of
information to test INSs' capability to accelerate particles.

\section*{Acknowledgments}
Tha analysis of XMM-Newton as well as Chandra data is supported
by the Italian Space Agency (ASI).

\begin{small}

\end{small}

\end{document}